\documentstyle[12pt]{article}

\begin{document}
\baselineskip=18pt plus 0.2pt minus 0.1pt
\parskip = 6pt plus 2pt minus 1pt
\newcommand{\reseteqnum}{\setcounter{equation}{0}}
\newcommand{\sD}{D\hspace{-0.63em}/}
\newcommand{\al}{\alpha}
\newcommand{\be}{\beta}
\newcommand{\ve}{\varepsilon}
\newcommand{\ep}{\epsilon}
\newcommand{\la}{\lambda}
\newcommand{\La}{\Lambda}
\newcommand{\ga}{\gamma}
\newcommand{\si}{\sigma}
\newcommand{\de}{\delta}
\newcommand{\bx}{\mbox{\boldmath $x$}}
\newcommand{\bX}{\mbox{\boldmath $X$}}
\newcommand{\bY}{\mbox{\boldmath $Y$}}
\newcommand{\D}{\partial}

\begin{titlepage}
\title{
\hfill
\parbox{4cm}{\normalsize KUNS-1349\\HE(TH)~95/09\\hep-ph/9508375}\\
\vspace{1cm}
Shapes of Cosmic Strings and Baryon Number Violation
}
\author{Masatoshi Sato\thanks{e-mail address:
\tt msato@gauge.scphys.kyoto-u.ac.jp}
{\,}\thanks{Fellow of the Japan Society for the Promotion of Science
for Japanese Junior Scientists.}
\\
{\normalsize\sl Department of Physics, Kyoto University}\\
{\normalsize\sl Kyoto 606-01, Japan}}
\date{\normalsize August, 1995}
\maketitle
\thispagestyle{empty}

\begin{abstract}
\normalsize
The relation between shapes of cosmic string and baryon number
violation is investigated.
If there exist fermionic zero modes on the string, using bosonization
technique, it is possible to obtain the effective action which describes
the fermion coupled to the arbitrarily shaped cosmic string.
The relation between baryon number and the sum of the writhing
number and linking number of the cosmic strings is rederived.
Baryons are created on the strings as the shapes of the cosmic strings change.
Furthermore we discuss implications of this baryon number violating
process to baryogenesis.
\end{abstract}
\end{titlepage}

\newpage
\reseteqnum
\section{Introduction}
Toward a better understanding of the physics on the string-like
objects, it is important to search for typical phenomena of their
structures of extent.
It is especially interesting to investigate the entanglement of the
string-like objects, which is inherent in the string-like objects in four
dimensional space-time.
Recently the present author and Yahikozawa attempted to find such phenomena
for the Nielsen-Olesen vortices \cite{SY}.
It was found that the expectation value of
$\int dx^{4} F_{\mu\nu}\tilde{F}_{\mu\nu}$ under the background of
the Nielsen-Olesen vortices becomes the difference between the sum of the
linking number and the writhing number of them at the initial time and
the one at the final time.
This means that if chiral fermions are coupled to the vortices,
fermion number is violated as vortices change their shapes, because of
the anomaly.
In this letter we include fermion in the system explicitly and closely
examine this mechanism of fermion number violation.

Another purpose of this letter is related to baryogenesis.
Baryogenesis has become a topic of much recent activity.
Not to contradict with inflation scenarios,
it is preferred that a baryon asymmetry is produced at lower
temperature.
In many GUT theories, the reheating temperature after inflation must be
lower than GUT scale \cite{Sat}-\cite{ELN}.
This infers that baryogenesis at GUT scale \cite{Yo}-\cite{sWa} is
incompatible with inflation.
To consider alternative scenarios of baryogenesis, we must introduce
a baryon number violating process at lower than GUT scale without
contradiction to proton decay experiments.
In most alternative baryogenesis scenarios, the EW sphaleron \cite{KM}
was used \cite{KRS,CKN,FY}.
Our mechanism of fermion number violation is another way to
introduce a baryon number violating process without contradiction to
proton decay experiments.
We also investigate implications of our fermion number violating
process to baryogenesis.

There are related papers to this work.
The relation between the linking number of the strings and baryon
number was discussed in \cite{CM,VF,GV}.
However, they neglected the contribution from the writhing number of
the strings.
Inclusion of fermion into the system was attempted in
\cite{GV}, but an only restricted configuration of the strings was
considered.

In the following, we will construct the effective action of the fermion
coupled to {\em arbitrarily shaped} strings and show explicitly that
baryons and leptons are created on the strings as they change their
shapes.
Furthermore we discuss its implications to baryogenesis.
Units in which $k_{{\rm B}}=\hbar=c=1$ will be used and $m_{{\rm Pl}}$ is the
Planck mass.

\section{The construction of effective action of fermion coupled to
arbitrarily shaped strings}
The model which we will consider is
\begin{eqnarray}
&&{\cal L}={\cal L}_{gauge-Higgs}+{\cal L}_{fermion},\nonumber\\
&&{\cal L}_{gauge-Higgs}=-\frac{1}{4}Z_{\mu\nu}Z^{\mu\nu}
                      +D_{\mu}\varphi^{*}D^{\mu}\varphi
                      -\la(|\varphi|^2-v^{2})^2,
                       \nonumber\\
&&{\cal L}_{fermion}=\bar{U}_{L}i\sD\, U_{L}+\bar{U}_{R}i\sD\, U_{R}
         -y_{q}(\bar{U}_{R}\varphi U_{L}+\,{\rm h.c.})
         \nonumber\\
&&\qquad\qquad+\bar{E}_{L}i\sD\,E_{L}+\bar{E}_{R}i\sD\,E_{R}
        -y_{l}(\bar{E}_{R}\varphi^{*}E_{L}+\,{\rm h.c.}),
\end{eqnarray}
where $Z_{\mu\nu}=\D_{\mu}Z_{\nu}-\D_{\nu}Z_{\mu}$ and
$D_{\mu}\varphi=(\D_{\mu}-iA_{\mu})\varphi$. The $U(1)$ charges
are assigned as follows.
\begin{center}
\begin{tabular}{c|c|c|c|c|c}\hline
 ${\rm matter}$
 & $U_{R}$ & $U_{L}$ & $E_{R}$ & $E_{L}$ & $\varphi$\\ \hline
 ${\rm charge}$
 & $r+1/2$    & $r-1/2$  & $\hat{r}-1/2$    & $\hat{r}+1/2$ & +1\\ \hline
\end{tabular}
\end{center}
\begin{center}
Table I. $\,U(1)$ charge assignment
\end{center}
The anomaly cancellation condition requires that $r$ and $\hat{r}$ must
satisfy the relation $r^{2}=\hat{r}^{2}$.
Because of the anomaly, the ``baryon'' number current $j_{B}^{\mu}\equiv
\bar{U}\gamma^{\mu}\,U$ and the ``lepton'' number current
$j_{L}^{\mu}\equiv\bar{E}\gamma^{\mu}\,E$ do not conserve and
satisfy the following relations.
\begin{eqnarray}
&&\D_{\mu}j^{\mu}_{B}=-\frac{r}{8\pi^{2}}
                 \tilde{Z}^{\mu\nu}Z_{\mu\nu},
\nonumber\\
&&\D_{\mu}j^{\mu}_{L}=\frac{\hat{r}}{8\pi^{2}}
                            \tilde{Z}^{\mu\nu}Z_{\mu\nu},
\label{anomaly:eqn}
\end{eqnarray}
where $\tilde{Z}^{\mu\nu}=\ep^{\mu\nu\rho\si}Z_{\rho\si}/2$ and
$\ep^{0123}=-\ep_{0123}=1$.
{}From these equation, we obtain
\begin{eqnarray}
&&Q_{B}=-\frac{r}{8\pi^{2}}\ep_{ijk}\int d^{3}xZ_{i}Z_{jk},\nonumber\\
&&Q_{L}=\frac{\hat{r}}{8\pi^{2}}\ep_{ijk}\int d^{3}xZ_{i}Z_{jk}.
\end{eqnarray}
When the $U(1)$ symmetry is spontaneously broken, cosmic strings form.
Under the background of the strings, the flux is concentrated on the
strings and this can be expressed as
\begin{equation}
Z_{\mu\nu}(x)=-\frac{1}{4}\ep_{\mu\nu\rho\si}J^{\rho\si}(x),
\end{equation}
where
\begin{equation}
J^{\mu\nu}(x)=4\pi\sum_{I}\int d^{2}\tau
             \frac{\D X_{I}^{[\mu}}{\D \tau^{0}}
             \frac{\D X_{I}^{\nu]}}{\D \tau^{1}}
             \de^{4}(x-X_{I}(\tau)).
\end{equation}
Here $X_{I}^{\mu}$ denotes the position of the I-th cosmic string and
$\tau$ is the coordinate parameterizing the world sheet swept by
the cosmic string.
Solving this equation under the gauge conditions
$\tau^{0}=X_{I}^{0}$ and $\D_{i}Z_{i}=0$, $Z_{\mu}$ become
\begin{eqnarray}
&&Z_{0}(\bx,t)=-\frac{1}{2}\ep_{ijk}\sum_{I}\oint d\tau^{1}
           \frac{\D X_{I}^{i}}{\D\tau^{1}}\frac{\D X_{I}^{j}}{\D t}
           \D_{k}\left(\frac{1}{|\bx-\bX_{I}|}\right),
\nonumber\\
&&Z_{i}(\bx,t)=-\frac{1}{2}\ep_{ijk}\sum_{I}\oint d\tau^{1}
           \frac{\D X_{I}^{j}}{\D\tau^{1}}
           \D_{k}\left(\frac{1}{|\bx-\bX_{I}|}\right).
\label{4gauge:eqn}
\end{eqnarray}
Using this and (\ref{anomaly:eqn}), we obtain
\begin{eqnarray}
&&Q_{B}=-r\left(\sum_{I}W_{r}(\bX_{I})
          +\sum_{I \neq J}L_{k}(\bX_{I},\bX_{J})\right),\nonumber\\
&&Q_{L}=\hat{r}\left(\sum_{I}W_{r}(\bX_{I})
                +\sum_{I \neq J}L_{k}(\bX_{I},\bX_{J})\right).
\label{4bary:eqn}
\end{eqnarray}
The definition of $W_{r}$ and $L_{k}$ are
\begin{eqnarray}
&&W_{r}(\bX)=\frac{1}{4\pi}\oint d\tau_{1}\oint d\tau'_{1}\ep_{ijk}
       \frac{\D X^{i}}{\D \tau_{1}} \frac{\D X^{j}}{\D \tau'_{1}}
       \frac{(X(\tau)-X(\tau'))^{k}}{|\bX(\tau)-\bX(\tau')|^{3}},
       \nonumber\\
&&L_{k}(\bX,\bY)
       =\frac{1}{4\pi}\oint d\tau_{1}\oint d\tau'_{1}\ep_{ijk}
       \frac{\D X^{i}}{\D \tau_{1}} \frac{\D Y^{j}}{\D \tau'_{1}}
       \frac{(X(\tau)-Y(\tau'))^{k}}
        {|\bX(\tau)-\bY(\tau')|^{3}}.
\label{wr,lk:eqn}
\end{eqnarray}
$L_{k}$ is the Gauss linking number and it takes an integer.
$W_{k}$ is called the writhing number and in general not takes
integer \cite{Ca}.
This is a geometrical quantity and measures the twist of the cosmic string.
Unless the cosmic string intersects with itself,
the value of the writhing number changes continuously as the shapes of
the string change.
When the string intersects with itself, the value of the writhing number
changes by 2.
An explicit example of the writhing number was given in \cite{FKV}.

{}From (\ref{4bary:eqn}), it is found that the ``baryon'' number and the
``lepton'' number change according to the shapes of cosmic strings.
To examine this more closely, we will analysis the fermionic
part of the system.

On the cosmic string, the vacuum expectation value of the Higgs field
$\varphi$ becomes zero.
Since the fermions get masses by the Yukawa coupling, massless
fermionic modes exist on the string.
In the following, we will derive the effective action of these
massless modes and show that these massless modes are produced
on the strings as ``baryon'' or ``lepton''.

First we will derive the effective action of the massless modes on the
z-directed cosmic string.
The field configuration of $\varphi$ and $Z_{\mu}$ are given as the
Nielsen-Olesen solution $\varphi^{\rm vortex}$ and $Z_{\mu}^{\rm
vortex}$ \cite{NO}.
On this background, it is known that
there exists a solution of the Dirac equation which is
independent on z and t and normalized in the x-y plane \cite{JR,We}.
We denote this as $(U^{\dot{\alpha}}_{L},U_{R \alpha},
E^{\dot{\alpha}}_{L},E_{R \alpha})
=(\be^{\dot{\alpha}}_{L}(x,y),\be_{R \alpha}(x,y),
\hat{\be}^{\dot{\alpha}}_{L}(x,y),\hat{\be}_{R \alpha}(x,y))$.
$\be$ and $\hat{\be}$ satisfy
$\si_{3}\be_{R}=\be_{R}$, $\si_{3}\be_{L}=-\be_{L}$,
$\si_{3}\hat{\be}_{R}=-\hat{\be}_{R}$ and
$\si_{3}\hat{\be}_{L}=\hat{\be}_{L}$.
We take the normalization as $\int dxdy |\be|^{2}=1/2$.
Using this solution we can express the fermion constrained
on the string as
\begin{equation}
\left\{
\begin{array}{l}
U^{\dot{\al}}_{L}(x,y,z,t)=q(z,t)\be^{\dot{\al}}_{L}(x,y)\\
U_{R \al}(x,y,z,t)=q(z,t)\be_{R \al}(x,y)
\end{array}
\right. ,
\end{equation}
\begin{equation}
\left\{
\begin{array}{l}
E^{\dot{\al}}_{L}(x,y,z,t)=l(z,t)\hat{\be}^{\dot{\al}}_{L}(x,y)\\
E_{R \al}(x,y,z,t)
=l(z,t)\hat{\be}_{R \al}(x,y)
\end{array}
\right. .
\end{equation}
Substituting these configuration for
${\cal L}_{fermion}$ and integrating over x and y, we obtain
\begin{equation}
L_{eff}=iq^{*}(\D_{0}-irz_{0})q+iq^{*}(\D_{3}-irz_{3})q
       +il^{*}(\D_{0}-i\hat{r}z_{0})l-il^{*}(\D_{3}-i\hat{r}z_{3})l,
\end{equation}
where we define $L_{eff}$ as $\int dxdy{\cal L}_{fermion}$.
The gauge fields $z_{a}$ are $\mu=0,3$ components of the
$Z_{\mu}^{\rm vortex}$.
For the z-directed string they can be zero, but for arbitrarily shaped
strings they are not zero.

If we use $q_{r}\equiv(q\,,\,0)^{T}$ and
$l_{l}\equiv(0\,,\,l)^{T}$, this effective lagrangian can be rewritten as the
following more familiar form
\begin{equation}
L_{eff} =i\bar{q}_{r}\ga^{a}\left(\D_{a}-irz_{a}\right)q_{r}
        +i\bar{l}_{l}\ga^{a}(\D_{a}-i\hat{r}z_{a})l_{l},
\end{equation}
where $a=0,3$ and the 2-dimensional gamma matrix is defined as
$\ga^{0}=\si^{1}$ and $\ga^{3}=-i\si^{2}$.

The ``baryon'' number current and the ``lepton'' number current on the
string can be obtained in the similar manner.
These are defined as
$J^{a}_{B}=\int dx dy j^{\mu=a}_{B}$ and $J^{a}_{L}=\int dx dy
j^{\mu=a}_{L}$  and become
\begin{eqnarray}
&&J^{a}_{B}=\bar{q}_{r}\ga^{a}q_{r},
\nonumber\\
&&J^{a}_{L}= \bar{l}_{l}\ga^{a}l_{l}.
\end{eqnarray}

To discuss the effective action on the arbitrarily shaped string,
bosonization technique is useful, since in the bosonized fields the anomaly can
be treated classically \cite{Wi,Nac}.
Bosonization rules are given by
$i\bar{q}\ga^{a}\D_{a}q
   \rightarrow \frac{1}{2}\D_{a}\phi_{q}\D^{a}\phi_{q}$,
$\bar{q}\ga^{a}q
   \rightarrow\frac{1}{\sqrt{\pi}}\ep^{ab}\D_{b}\phi_{q}$ \cite{Co,Ma}.
Using
the relation $\bar{q}\ga^{a}\ga_{5}q=-\ep^{ab}\bar{q}\ga_{b}q$, we obtain
\begin{eqnarray}
L_{eff}&=&\frac{1}{2}\D_{a}\phi_{q}\D^{a}\phi_{q}
  +\frac{r}{2\sqrt{\pi}}(\ep^{ab}-\eta^{ab})z_{a}\D_{b}\phi_{q}
  +\frac{r^{2}}{8\pi}z_{a}z^{a}\nonumber\\
  &+&\frac{1}{2}\D_{a}\phi_{l}\D^{a}\phi_{l}
  +\frac{\hat{r}}{2\sqrt{\pi}}(\ep^{ab}+\eta^{ab})z_{a}\D_{b}\phi_{l}
  +\frac{\hat{r}^{2}}{8\pi}z_{a}z^{a},
\end{eqnarray}
and
\begin{eqnarray}
&&J_{B}^{a}=\frac{1}{2\sqrt{\pi}}(\ep^{ab}-\eta^{ab})\D_{b}\phi_{q}
         -\frac{r}{4\pi}(\ep^{ab}-\eta^{ab})z_{b},
\nonumber\\
&&J_{L}^{a}=\frac{1}{2\sqrt{\pi}}(\ep^{ab}+\eta^{ab})\D_{b}\phi_{l}
         +\frac{\hat{r}}{4\pi}(\ep^{ab}+\eta^{ab})z_{b}.
\end{eqnarray}
Here $\phi_{q}$ and $\phi_{l}$ are bosonized fields of $q$ and $l$
respectively.
The terms which do not depend on $\phi_{q}$ or $\phi_{l}$ are
determined by gauge invariance.
For the bosonization fields, gauge transformation can be defined as
\begin{eqnarray}
&&z_{a}\rightarrow z_{a}+\D_{a}\La,\nonumber\\
&&\phi_{q}\rightarrow \phi_{q}+\frac{r\La}{2\sqrt{\pi}},\nonumber\\
&&\phi_{l}\rightarrow \phi_{l}-\frac{\hat{r}\La}{2\sqrt{\pi}}.
\end{eqnarray}
This gauge transformation is expected from the operator relation of
$q$ and $\phi_{q}$, and only when the anomaly cancellation condition is
satisfied the bosonized action becomes gauge invariant.

Now we will consider the effective action of the massless modes on an
arbitrarily shaped string $X^{\mu}(\tau)$.
This action is obtained from the above action by the replacement
$\eta^{ab}\rightarrow \sqrt{g}g^{ab}$ and $(z,t)\rightarrow
(\tau^{0},\tau^{1})$, where $g_{ab}=\D_{a}X^{\mu}\D_{b}X_{\mu}$ is
the induced metric on the string and $g=\det g_{ab}$.
The effective action is
\begin{eqnarray}
L_{eff}&=&\frac{1}{2}\sqrt{-g}g^{ab}\D_{a}\phi_{q}\D_{b}\phi_{q}
  +\frac{r}{2\sqrt{\pi}}(\ep^{ab}-\sqrt{-g}g^{ab})z_{a}\D_{b}\phi_{q}
  +\frac{r^{2}}{8\pi}\sqrt{-g}g^{ab}z_{a}z_{b}\nonumber\\
&+&\frac{1}{2}\sqrt{-g}g^{ab}\D_{a}\phi_{l}\D_{b}\phi_{l}
  +\frac{\hat{r}}{2\sqrt{\pi}}(\ep^{ab}+\sqrt{-g}g^{ab})z_{a}\D_{b}\phi_{q}
  +\frac{\hat{r}^{2}}{8\pi}\sqrt{-g}g^{ab}z_{a}z_{b}
\label{ef-ac:eqn}
\end{eqnarray}
and the ``baryon'' number current and the ``lepton'' number current on the
arbitrarily shaped string are
\begin{eqnarray}
&&\sqrt{-g}J_{B}^{a}
         =\frac{1}{2\sqrt{\pi}}(\ep^{ab}-\sqrt{-g}g^{ab})\D_{b}\phi_{q}
         -\frac{r}{4\pi}(\ep^{ab}-\sqrt{-g}g^{ab})z_{b},
\nonumber\\
&&\sqrt{-g}J_{L}^{a}
         =\frac{1}{2\sqrt{\pi}}(\ep^{ab}+\sqrt{-g}g^{ab})\D_{b}\phi_{l}
         +\frac{\hat{r}}{4\pi}(\ep^{ab}+\sqrt{-g}g^{ab})z_{b}.
\end{eqnarray}

Considering the massless modes on the I-th cosmic string, $z_{a}(\tau)$ are
given by $z_{a}(\tau)=(\D X_{I}^{\mu}/\D \tau^{a})Z_{\mu}^{\rm
vortex}(X(\tau))$.
When the radius of curvature of the string is much greater than the
string thickness, $Z_{\mu}^{\rm vortex}$ are given as
(\ref{4gauge:eqn}), therefore
\begin{eqnarray}
&&z_{0}(\tau)=-\frac{1}{2}\ep_{ijk}\sum_{J}\oint d\tau'
               \frac{\D X_{J}^{i}}{\D\tau'}
           \left(\frac{\D X_{I}^{j}(\tau)}{\D t}
           -\frac{\D X_{J}^{j}(\tau')}{\D t}
           \right)
            \frac{(X_{I}(\tau)-X_{J}(\tau'))^{k}}
            {|\bX_{I}(\tau)-\bX_{J}(\tau')|^{3}},
\nonumber\\
&&z_{1}(\tau)=\frac{1}{2}\ep_{ijk}\sum_{J}\oint d\tau'
              \frac{\D X_{I}^{i}}{\D\tau}
              \frac{\D X_{J}^{j}}{\D\tau'}
             \frac{(X_{I}(\tau)-X_{J}(\tau'))^{k}}
              {|\bX_{I}(\tau)-\bX_{J}(\tau')|^{3}}.
\label{2gauge:eqn}
\end{eqnarray}
Note that $z_{a}(\tau)$ are not singular at $\tau=\tau'$. As far as
the string does not intersect, $z_{a}(\tau)$ are well-defined functions.
This non-singularity of $z_{a}$ is not depend on the gauge condition
we take.
Even if we use $\D_{\mu}Z^{\mu}=0$ gauge, they are not singular.
This enables us to tell what occurs on the string definitely and makes
the effective action (\ref{ef-ac:eqn}) useful.

Using the equation of motion, we obtain
\begin{eqnarray}
&&\D_{a}\sqrt{-g}J^{a}_{B}=-\frac{r}{4\pi}\ep^{ab}z_{ab},
\nonumber\\
&&\D_{a}\sqrt{-g}J^{a}_{L}=\frac{\hat{r}}{4\pi}\ep^{ab}z_{ab}.
\end{eqnarray}
{}From (\ref{2gauge:eqn}), we find that the ``baryon'' number on the string
$Q^{(2)}_{B}\equiv\oint d\tau^{1}\sqrt{-g}J^{a}_{B}$ and the
``lepton'' number on the string $Q^{(2)}_{L}\equiv\oint
d\tau^{1}\sqrt{-g}J^{a}_{L}$ satisfy
\begin{eqnarray}
&&\Delta Q^{(2)}_{B}=-r\Delta\left(W_{r}(\bX_{I})
                     +\sum_{J}L_{k}(\bX_{I},\bX_{J})\right),\nonumber\\
&&\Delta Q^{(2)}_{L}=\hat{r}\Delta\left(W_{r}(\bX_{I})
                     +\sum_{J}L_{k}(\bX_{I},\bX_{J})\right).
\label{2bary:eqn}
\end{eqnarray}
These relations are consistent with (\ref{4bary:eqn}) and show that
``baryons'' and ``leptons'' are produced on the strings according to
the changes of their shapes.

Although we have treated the only toy model in this section, the
conclusions do not change for more complicated models.
If all the fermions get masses by Yukawa coupling,
the effective action of the massless modes can be constructed in the
same manner and it is found that baryons and leptons are produced on the
string as the cosmic strings change their shapes.
Indeed in the cases of the electroweak string \cite{Na,Va} and cosmic
strings in supersymmetric model with an extra $U(1)$ \cite{TDB}, we have
constructed the effective actions of the string-fermion system except
the neutrino part \cite{SY2}.
If massless fermions exist, the above analysis do not apply in
general, however, since (\ref{4bary:eqn}) is irrelevant to the fermion
mass, we believe that the conclusions do not change in this case.

\section{Implications to baryogenesis}
In the previous section, we have shown that as cosmic strings
change their shapes, baryons are generated on them.
In the following we will discuss baryogenesis due to this
baryon number violation mechanism.

We will consider cosmic strings on which baryons are generated
by our mechanism of baryon number violation.
For example, the electroweak string is this type of the cosmic
strings.
Another example is cosmic strings in a supersymmeric model with an
extra $U(1)$.

As first pointed out by Sakharov, three conditions need to be
satisfied in order to generate a net baryon number \cite{Sa}.
First, baryon number violating process must exist.
Second, these process must violate C and CP.
And third, they must occur out of thermal equilibrium.
C and CP are violated in usual particle physics models
and once cosmic strings form, they become rapidly out of equilibrium.
Therefore we assume that second and third conditions are satisfied.
In the following we will concentrate on the first condition.

It is expected that our mechanism of baryon number violation is the most
efficient at string formation.
If cosmic strings form by the Kibble mechanism \cite{Ki}, the
correlation length of the strings $\xi$ at formation is given by
$\xi(\eta)\sim (\lambda\eta)^{-1}$, where $\eta$ is the scale of the
symmetry breaking by which cosmic string forms and $\lambda$ is a
Higgs self coupling constant or a gauge coupling constant.
Since the strings form by randomly assigning values of
the phase of the Higgs field,
it is expected that the string twists once per
$(\lambda\eta)^{-3}$ volumes.
Therefore the number of baryons or anti-baryons created
on the strings is about $(\lambda\eta)^{3}$ particles
per unit volume.
This is comparable to the entropy of the massless particles at $T=\eta$.

Although there exists a baryon number violating process, it is
not obvious to generate a net baryon number because the baryon number
is violated due to the anomaly.
In this case we find that a linear combination of the
baryon number and the Chern-Simons form preserves.
This gives a restriction to generate a net baryon number.
For example, in our toy model,
\begin{equation}
Q_{B}+\frac{r}{8\pi^{2}}\epsilon_{ijk}\int dx^{3}Z_{i}Z_{jk}
\label{eqn:bn}
\end{equation}
preserves.
This tells us that
if there is no baryon before cosmic strings form, no net
baryon remains after cosmic strings vanish.
This is because in the toy model cosmic strings forms by spontaneous
breaking of $U(1)$ gauge symmetry.
Since $U(1)$ gauge symmetry has only one vacuum, no net baryon number
is produced by the anomaly after cosmic strings vanish.

One possibility to obtain a net baryon number is to combine with
an extra baryon number violating process.
A candidate of the extra baryon number violating process is the
sphaleron (or instanton) transition.
An explicit baryon number violating process in GUT is another
candidate, but if we consider baryogenesis scenarios after
inflation, the sphaleron transition is more attractive since it does not
contradict with experiments of the proton decay.

We assume that the extra baryon number violating
process exists and the rate of the process is high at string formation.
We further assume that after $T=\eta'$, this process becomes
suppressed because of a phase transition.
Until $T=\eta'$, the baryon number produced by the strings is erased
by the extra baryon number violating process.
Thus $Q_{B}$ in (\ref{eqn:bn}) is zero at $T=\eta'$.
After $T=\eta'$, the linear combination of the baryon number and the
Chern-Simon form (\ref{eqn:bn}) preserves, because the extra baryon
number violating process becomes irrelevant.
Therefore we find that a net baryon
number remains after the strings vanish and this is given by the sum
of the writhing number and the linking number of the strings at
$T=\eta'$.

To estimate the sum of the writhing number at $T=\eta'$,
considerations about the evolution of the string is needed.
After formation, strings experience a significant damping force from
the high plasma background density.
A heuristic argument \cite{Ki} show that in the friction dominated
epoch the correlation length $\xi$ change as
\begin{equation}
|\,\xi(T)-\xi(\eta)\,|
\sim\left(\frac{m_{{\rm Pl}}(\eta^2-T^2)}{T^5}\right)^{1/2}.
\label{eqn:xi}
\end{equation}
This infers that the structures of the string smaller than
$(m_{{\rm Pl}}/\eta^{3})^{1/2}$ are straightened when $T$ becomes
slightly lower than $\eta$.
This result is confirmed by solving the equation of motion.
We take the metric of the Friedman universe as
$ds^{2}=a^{2}(\tau)(d\tau^{2}-d\bx^{2})$, where $\tau$ is conformal
time $d\tau=dt/a(t)$.
In the gauge conditions $\tau_{0}=\tau$ and $\dot{\bx}\cdot\bx'=0$,
the string evolves as
\begin{equation}
\ddot{\bx}-\epsilon^{-1}\left(\frac{\bx'}{\epsilon}\right)'
+\left(2\frac{\dot{a}}{a}+\frac{\beta T^{3}}{\mu}a\right)
(1-\dot{\bx}^{2})\dot{\bx}=0,
\end{equation}
where $\epsilon$ is given by
\begin{equation}
\epsilon=\left(\frac{\bx'^{2}}{1-\dot{\bx}^{2}}\right)^{1/2}
\end{equation}
and $\mu\equiv\eta^{2}$ is the string tension and $\beta$ is
numerical constant of order unity \cite{Vi}.
Here dots and primes denote derivatives with respect $\tau_{0}$ and
$\tau_{1}$.
Let us examine small perturbations on the
static straight string.
If we represent $\bx$ as
$\bx=\mbox{\boldmath $e$}\tau_{1}+\delta\bx(\tau_{0})e^{ik\tau_{1}}$,
where $\mbox{\boldmath $e$}$ is an unit vector, and neglect the Hubble
damping term,
the equation of motion becomes
\begin{equation}
\delta\ddot{\bx}
+\frac{\beta T^{3}}{\mu}a
\delta\dot{\bx}+k^{2}\delta{\bx}=0.
\end{equation}
In the radiation era, $a$ and $T$ are given as $a\sim m_{{\rm Pl}}\tau_{0}$
and $T\sim 2/\tau_{0}$.
In order to examine this equation at $\tau_{0}\sim\eta^{-1}$, it is sufficient
to solve the equation where the coefficient of $\delta\dot{\bx}$ is
replaced with a constant $\beta m_{{\rm Pl}}$.
The solution can be easily obtained and we find that if $k/a$ is
larger than $(\eta^{3}/m_{{\rm Pl}})^{1/2}$, the amplitude of $\delta\bx$ is
damped within $\Delta\tau_{0}\sim 1/\eta$.
This means that structures of the strings smaller than
$(m_{{\rm Pl}}/\eta^{3})^{1/2}$ are straightened within $\Delta T\sim\eta'$.
This result agrees with the heuristic argument\footnote{
Same conclusion was obtained in \cite{GS,MS}.}.

We have found that a net baryon obtained is
determined by the sum of writhing number and the linking number of the
strings at $T=\eta'$.
The sum of the writhing number per unit volume is expected to be
$\xi(\eta')^{-3}$.
If CP violation bias parameter is denoted as $\varepsilon$, we obtain
a net baryon number per unit volume as $\varepsilon\,\xi(\eta')^{-3}$.
Since the entropy of massless particles is given by $s\sim T^{3}$,
the baryon to entropy ration will be
\begin{equation}
\frac{n_{_B}}{s}\sim\varepsilon
\left(\frac{\eta'^3}{m_{{\rm Pl}}(\eta^2-\eta'^2)}\right)^{3/2}.
\end{equation}
This gives a severe constraint to work the scenario.
Unless $\eta$ is close to $m_{{\rm Pl}}$ or $\varepsilon$ is very large, we
find that $\eta'$ must almost coincide with $\eta$ to obtain a baryon
asymmetry of the magnitude required to explain the present baryon to
entropy ratio.
If we consider cosmic string forming at $O({\rm TeV})$ and the EW sphaleron
as the extra baryon number violating process, it is impossible to obtain
the baryon asymmetry.

There is another possibility to obtain a net baryon number.
Let us consider the string-forming phase transition $G\rightarrow H$.
If $\pi_{1}(G/H)$ is non-abelian, a topological
argument tells that the strings corresponding to non-commuting elements of
$\pi_1$  cannot pass through each other \cite{Ki2,VS}.
Therefore the linking number of the cosmic strings cannot almost
change and a net baryon number remains.
Vachaspati and Field estimated numerically that the linking number of
the cosmic strings per unit volume is $10^{-4}\,(\lambda\eta)^{-3}$ \cite{VF}.
This gives the baryon to entropy ratio as $10^{-4}\varepsilon\lambda^{-3}$.

We comment the evolution of this type of strings.
It was pointed out that this type of cosmic strings form a tangled
network and if they form by the Kibble mechanism, they
dominate the universe very soon after formation \cite{Vi2}.
This is an obstacle to the baryogenesis scenario.
However, the conclusion changes if we take into account
intercommuting of the string at crossing itself.
Since an element of $\pi_1$ commutes with itself, the topological
argument gives no constraint when the string crosses itself.
Then this intercommuting is possible.
This enables not to form a tangled string network, since by
intercommuting, tangled long strings become loosely linked long
strings with many linked tiny strings.
Therefore too early string-domination in universe does not occur.

\section{Conclusions}
In this letter we have presented the effective action of fermion
coupled to arbitrarily shaped strings.
As was suggested in \cite{SY}, the fermion are created as the strings
change their shapes.
Furthermore we have discussed its implication to baryogenesis.
We have considered two possibilities to obtain a net baryon number.
The first is to combine with an extra baryon number
violating process.
Considered carefully the evolution of cosmic strings in
the friction dominated epoch, it has been shown that string-forming
scale must be very close to the scale where the extra baryon number
violating process begins to suppress.
This is because the structures of the strings smaller than
$(m_{{\rm Pl}}/\eta^{3})^{1/2}$ are straighten within $\Delta T\sim\eta$.
This constraint is not inherent in our baryon number violating
process.
Defect-mediated electroweak baryogenesis \cite{BDT,TDB} also suffers
this constraint
\footnote{Eq.(3.10) in \cite{BDT} is not correct and their conclusion
must be changed.}.
As a second possibility, we have considered the cosmic string with
non-abelian $\pi_{1}(G/H)$.
It is possible to produce a baryon asymmetry without too
early string-domination in universe.

\vskip1cm
\centerline{\large\bf Acknowledgments}
\noindent
The author would like to thank S. Yahikozawa for reading the manuscript and
useful comments.

\newpage

\newcommand{\J}[4]{{\sl #1} {\bf #2} (19#3) #4}
\newcommand{\MPL}{Mod.~Phys.~Lett.}
\newcommand{\NP}{Nucl.~Phys.}
\newcommand{\PL}{Phys.~Lett.}
\newcommand{\PR}{Phys.~Rev.}
\newcommand{\PRL}{Phys.~Rev.~Lett.}
\newcommand{\AP}{Ann.~Phys.}
\newcommand{\CMP}{Commun.~Math.~Phys.}
\newcommand{\CQG}{Class.~Quant.~Grav.}
\newcommand{\PRP}{Phys.~Rept.}
\newcommand{\SPU}{Sov.~Phys.~Usp.}
\newcommand{\RMPA}{Rev.~Math.~Pur.~et~Appl.}
\newcommand{\SPJ}{Sov.~Phys.~JETP}
\newcommand{\UFN}{Usp.~Fiz.~Nauk}
\newcommand{\JP}{J.~Phys.}
\newcommand{\JETPL}{JETP~Letters}
\newcommand{\PZETF}{Pis'ma~Zh.~Eksp.~Teor.~Fiz.}
\newcommand{\MNRAS}{Mon.~Not.~R.~Astron.~Soc.}
\newcommand{\ARNPS}{Ann.~Rev.~Nucl.~Part.~Sci.}
\newcommand{\APP}{Acta~Phys.~Polon.}

\end{document}